\RequirePackage{fix-cm}
\documentclass[twocolumn,epjc3]{svjour3}  
\smartqed  
\RequirePackage{graphicx}
\usepackage{graphics}
\usepackage{multicol}
\usepackage{graphicx}
\usepackage{booktabs}
\usepackage{amssymb,bm,mathrsfs,bbm,amscd}
\usepackage[tbtags]{amsmath}
\usepackage{lastpage}
\usepackage{changes}
\usepackage{dcolumn}
\usepackage{longtable}
\usepackage[numbers,sort&compress]{natbib}
\usepackage{graphicx}
\usepackage{dcolumn}
\usepackage{bm}
\usepackage{amsmath}
\usepackage{flushend}
\usepackage{changes}
\usepackage[colorlinks,linkcolor=blue,anchorcolor=blue,citecolor=blue,urlcolor=blue]{hyperref}
\usepackage{threeparttable}
\usepackage{subfigure}

\usepackage{array}
\newcommand{\PreserveBackslash}[1]{\let\temp=\\#1\let\\=\temp}
\newcolumntype{C}[1]{>{\PreserveBackslash\centering}p{#1}}
\newcolumntype{R}[1]{>{\PreserveBackslash\raggedleft}p{#1}}
\newcolumntype{L}[1]{>{\PreserveBackslash\raggedright}p{#1}}

\journalname{Eur. Phys. J. A}
\begin{document}
\title{Systematic study on proton radioactivity of spherical proton emitters within two--potential approach}
\author{Jiu-Long Chen\thanksref{addr1}
\and Xiao-Hua Li\thanksref{addr1,addr2,addr3,e1}
\and Xi-Jun Wu\thanksref{addr4,e2}
\and Peng-Cheng Chu\thanksref{addr5,e3}
\and Biao He\thanksref{addr6}%
}
\thankstext{e1}{e-mail: lixiaohuaphysics@126.com}
\thankstext{e2}{e-mail: wuxijun1980@yahoo.cn}
\thankstext{e3}{e-mail: kyois@126.com}


\institute{School of Nuclear Science and Technology, University of South China, Hengyang 421001, China \label{addr1}
\and Cooperative Innovation Center for Nuclear Fuel Cycle Technology $\&$ Equipment, University of South China, Hengyang 421001, China \label{addr2}
\and Key Laboratory of Low Dimensional Quantum Structures and Quantum Control, Hunan Normal University, Changsha 410081, China \label{addr3}
\and School of Math and Physics, University of South China, Hengyang 421001, China\label{addr4}
\and School of Science, Qingdao Technological University, Qingdao 266000, China\label{addr5}
\and College of Physics and Electronics, Central South University, Changsha 410083, China \label{addr6}}
\date{Received: date / Accepted: date}

\maketitle
\begin{abstract}
In the present work we systematically study the half--lives of proton radioactivity for spherical proton emitters with ${Z\ge 69}$ based on two--potential approach. While the nuclear potential of the emitted proton--daughter nucleus is adopted by a parameterized cosh type, the parameters of the depth and diffuseness for nuclear potential are determined by fitting experimental data of 32 spherical proton emitters. In order  to reduce the deviations between experimental half-lives and calculated ones, we propose a simple analytic expression for formation probability of proton radioactivity with the same orbital angular momentum $l$. The results indicate that the formation probability can be simply described by a formula of $A_d^{1/3}$. Moreover, the linear relationship between the formation probability and the fragmentation potential also exists. The calculated half-lives can well reproduce the experimental data.
\end{abstract}
%

\section{Introduction}
\label{section 1}

Proton radioactivity is an important decay mode of proton-rich nuclide far away from $\beta-$stability line. Study on this decay process can well promote the existing nuclear theories and models, and also drive the development of more models. The first clear evidence for proton emission with a measurable half-life was obtained in the studies on the decay of an isomeric state for $^{53}$Co \cite{JACKSON1970281,CERNY1970284}.  In 1981, Hofmann ${\it et\  al.}$ \cite{Hofmann1982}detected the proton emission from nuclear ground state of $^{151}$Lu at the SHIP separator of GSI. Klepper ${\it et\  al.}$ \cite{Klepper1982} and  Faestermann ${\it et\  al.}$\cite{Faestermann1984}  observed proton emission from the ground state of $^{147}$Tm and $^{109}$I, $^{113}$Cs in 1982 and 1984, respectively. With the development of advanced experimental facilities and radioactive beams, an increasing number of proton emissions from the ground state or low isomeric states have been discovered in the proton regions $Z=50-82$ \cite{SONZOGNI20021,PhysRevLett.96.072501,BLANK2008403,Zhang_2010,
PhysRevC.96.034619,Budaca2017,1674-1137-41-3-030001,Chen_2019}. As an important decay mode of unstable nuclei, proton radioactivity is an useful tool to obtain spectroscopic information because the decaying proton is the unpaired proton not filling its orbit, and extract information on nuclear structure and the internuclear potential \cite{KARNY200852}. 

The proton radioactivity has the lowest Coulomb potential among all charged particles and mass being smallest it suffers the highest centrifugal barrier, enabling this process suitable to be dealt within WKB barrier penetration model \cite{PhysRevC.72.051601}. Up to now, there are a lot of models having been put forward to deal with the proton radioactivity such as the effective interactions of density--dependent M3Y (DDM3Y) \cite{BHATTACHARYA2007263,QIAN-Yi-Bin-72301}, the single--folding model (SFM) \cite{PhysRevC.72.051601,QIAN-Yi-Bin**-112301}, the generalized liquid--drop model (GLDM) \cite{PhysRevC.79.054330,Zhang_2010,yzwang2017},  the phenomenological unified fission model (UFM) \cite{PhysRevC.71.014603,1674-1137-34-2-005}, the Coulomb and proximity potential model (CPPM) \cite{PhysRevC.96.034619,ZhangGL2013,ZhangGL2014}, the Gamow-like model (GLM) \cite{zdeb2016,Chen_2019}, the two--potential approach with Skyrme-Hartree-Fock (TPA-SHF) \cite{CHENG2020121717} and so on. As we all known, in the process of studying the charged particles radioactivity, selection of the emitted particle-nucleus interaction potential is key to improve the accuracy of half-life. In 1992, Buck ${\it et\  al.}$ \cite{PhysRevC.45.2247,PhysRevLett.72.1326} proposed a cluster model to study $\alpha$ decay. Their calculated results can successfully reproduce the experimental data of $\alpha$ decay half--lives. While the nuclear potential of the emitted $\alpha$--daughter nucleus is adopted by a parameterized cosh type. In our previous works \cite{PhysRevC.93.034316,PhysRevC.94.024338,PhysRevC.95.044303,Deng_2018}, we used the two--potential approach (TPA) \cite{PhysRevLett.59.262} with cosh type potential to systematically study  $\alpha$ decay half-lives and the calculations can reproduce the experimental data well. Zhang ${\it et\  al.}$ \cite{Zhang072301} improved the cluster model with cosh type potential to explain the proton radioactivity and proposed a phenomenological formula to calculate the half--life of this process. In this work, based on TPA,  we systematically study the proton radioactivity of spherical proton emitters. While the total emitted proton-daughter nucleus potential $V(r)$ is composed of the cosh type nuclear potential $V_N(r)$, Coulomb potential $V_C(r)$ and centrifugal potential $V_l(r)$. The calculated results adopted the TPA with cosh type potential can reproduce the experimental data well.

Moreover, since the formation probability (spectroscopic factor) of proton radioactivity being one of the most important considerations in the calculation of half-life, it has becoming a hot topic \cite{PhysRevC.56.1762, DELION2006113,Qian201668,Soylu_2021,PhysRevC.90.054326,PhysRevC.93.014314}. There are several semi-microscopic methods to computationally study the formation probability such as relativistic mean field theory (RMF) with BCS method \cite{Zhang_2010,PhysRevC.56.1762, DELION2006113,PhysRevC.79.054330,Qian201668,Soylu_2021}, covariant density functional (CDF) with BCS method \cite{PhysRevC.90.054326} and relativistic continuum Hartree-Bogoliubov (RCHB) \cite{PhysRevC.93.014314} and so on. While the phenomenological methods in relation to nuclear structure are also adopted to study the formation probability such as a phenomenological discussion about its linear relationship given by Qi ${\it et\  al.}$ \cite{PhysRevC.85.011303,QI2019214,QI2021136373}. Recently,  Delion ${\it et\  al.}$ \cite{PhysRevC.103.054325} discover a linear relationship between formation probability of proton radioactivity and $A^{1/3}$. In this sense, based on universal decay law of proton radioactivity (UDLP) and the phenomenological linear relationship of formation probability  proposed by Qi ${\it et\  al.}$ \cite{PhysRevC.85.011303}, we systematically study the relationship between the formation probability and the mass number of daughter nucleus $A_d$. The result indicates that the formation probability can be simply described by a formula of $A_d^{1/3}$. 

This article is organized as follows. In Section \ref{section 2}, the theoretical framework for calculation of proton radioactivity half--life is briefly described. The detailed calculations and discussion are presented in Section \ref{section 3}. Finally, a brief summary is given in Section \ref{section 4}.

\section{The theoretical method}
\label{section 2}
The half-life of proton radioactivity $T_{1/2}$ can be calculated as
\begin{equation}
T_{1/2}=\frac{\hbar\ln2}{\Gamma},
\end{equation}
where $\hbar$ is the reduced Planck constant,  $\Gamma$ is proton radioactivity width depending on the proton formation probability, the penetration probability and the collision probability. Within the TPA and the quasiclassical method \cite{PhysRevLett.59.262}, the width of the quasibound state in semiclassical approximation can be obtained expediently by the following as
\begin{equation}
\Gamma=S_{p}\frac{\hbar^{2}FP}{4\mu},
\end{equation}
where $\mu=m_pm_d/(m_p+m_d)\approx m_pA_d/(A_p+A_d)$ is the reduced mass of the decaying nuclear system with $m_p$ and $m_d$ being the mass of proton and daughter nucleus, respectively. $A_d$ and $A_p$ are the mass numbers of daughter nucleus and emitted proton. $S_p$ is the formation probability of the proton radioactivity. 

The normalized factor $F$, describing the assault frequency, is given by the integration over the internal region. The penetration probability $P$ is calculated by the semi--classical Wentzel-Kramers-Brilloum (WKB) approximation. In this study, the normalized factor $F$ and penetration probability $P$  describe only the standard semi-classical spherical approach. $F$ can be written as
\begin{equation}
F\int_{r_1}^{r_2}\frac{1}{2k(r)}dr=1,
\end{equation}
and $P$ can be expressed as
 \begin{equation}
P=\exp\left(-2\int_{r_2}^{r_3}k(r)dr\right),
 \end{equation}
where $k(r)=\sqrt{\frac{2\mu}{\hbar^2}|Q_p-V(r)|}$ is the wave number of emitted proton, $r$ is the distance between the emitted proton and the mass center of daughter nucleus. $Q_p$ is the released energy of proton radioactivity. $V(r)$ is the total emitted  proton-daughter nucleus interaction potential.  where $r_1$, $r_2$ and $r_3$ denote the classical turning points. They satisfy the conditions $V(r_i)=Q_p$ ($i=$1, 2, 3). The released energy $Q_p$  is generically calculated by \cite{PhysRevC.96.034619}
\begin{equation}\label{q-p}
Q_p=\Delta M-(\Delta M_d+\Delta M_p)+k(Z^{\varepsilon}-Z^{\varepsilon}_d),
\end{equation}
where $Z$ and $Z_d$ are the proton numbers of parent nucleus and daughter nucleus, respectively. The $\Delta M$, $\Delta M_d$ and $\Delta M_{p}$ are, correspondingly, the mass excess of parent and daughter nuclei and emitted proton. The experimental data of mass excess $\Delta M$, $\Delta M_d$ and $\Delta M_p$ are taken from the atomic mass table NUBASE2016 \cite{Huang_2017,Wang_2017}. The term $k(Z^{\varepsilon}-Z^{\varepsilon}_d)$ represents the screening effect of atomic electrons\cite{PhysRevC.72.064613}, where $k = 8.7$ eV, $\varepsilon = 2.517$ for $Z \ge 60$, and $k = 13.6$ eV, $\varepsilon = 2.408$ for $Z < 60$ \cite{HUANG1976243}.

The total emitted proton-daughter nucleus interaction potential $V(r)$ is composed of nuclear potential $V_N(r)$, Coulomb potential $V_C(r)$ and centrifugal potential $V_l(r)$. It can be expressed as
 \begin{equation}\label{vn+c+l}
V(r)=\lambda V_{N}(r)+V_{C}(r)+V_{l}(r),
 \end{equation}
where $\lambda$ is the renormalized factor. In the study of proton radioactivity, the total potential $V(r)$ within the TPA framework is split into two parts of internal bound state and external scattering state. The first one is given by Eq. \ref{vn+c+l} modified by a constant value beyond the barrier radius within the effect of $V_N(r)$, while the second one is entirely described by Eq. \ref{vn+c+l} without the effect of $V_N(r)$.  Bohr–Sommerfeld quantization condition is an important part of the WKB calculations that should be taken into account to meet the self-consistency \cite{Kelkar064605}. The renormalized factor $\lambda$  in the Eq. \ref{vn+c+l} is determined separately for each decay by employing this condition to generate a state of relative motion with $n$ nodes and orbital angular momentum $l$ at the released energy $Q_p$. It can be expressed as
\begin{equation}
\int_{r_1}^{r_2}\sqrt{\frac{2\mu}{\hbar^2}[Q_p-V(r)]}dr=(G-l+1)\frac{\pi}{2},
\end{equation}
where $G=2n+l$ is the principal quantum number. For proton radioactivity we choose G = 4 or 5 corresponding to the $4\hbar \omega$ or $5\hbar \omega$ oscillator shell depending on the individual proton emitter.

 In our previous works \cite{PhysRevC.93.034316,PhysRevC.94.024338,PhysRevC.95.044303,Deng_2018}, we chose nuclear potential $V_N(r)$ as  a type of cosh parameterized form proposed by Buck ${et\ al.}$ \cite{PhysRevC.45.2247} to systematically study the $\alpha$ decay half-lives of even-even, odd-odd and odd-$A$ nuclei. In this work, we generalize this cosh type nuclear potential to study proton radioactivity. It can be expressed as
 \begin{equation}\label{V_N}
V_{N}(r)=-V_0\frac{1+\rm{cosh}\it{(R/a)}}{\rm{cosh}\it{(r/a)}+\rm{cosh}\it{(R/a)}},
 \end{equation}
 where $V_0$ and $a$ are the parameters of the depth and diffuseness of nuclear potential, respectively. The Coulomb potential $V_C(r)$ is obtained under the assumption of a uniformly charged sphere. It can be written as
\begin{equation}
V(r)=
\begin{cases}
\frac{Z_de^2}{2R}\left[3-\frac{r^2}{R^2}\right]&\rm{for}\  r  \leq R,\\
\frac{Z_de^2}{r}&\rm{for}\  r > R,
\end{cases}
\end{equation}
where $R$ is the sharp radius. In this work, we use a semi-empirical formula in terms of mass number as $R=r_0A^{1/3}$ with $r_0=1.20$ fm \cite{Buck1688}. $A$ is the mass number of parent nucleus.

Because $l(l+1)\to (l+\frac{1}{2})^2$ is a necessary correction for one--dimensional problems \cite{doi:10.1063/1.531270}, we adopt the Langer modified centrifugal barrier. It can be written as
\begin{equation}
V_l(r)=\frac{\hbar^2(l+\frac{1}{2})^2}{2 {\mu}r^2},
\end{equation}
where  $l$ is the orbital angular momentum taken away by the emitted proton. The values of $l$ can be obtained by the  parity and angular momentum conservation laws as
\begin{equation}
J=J_d+J_{p}+l, \ \  \pi=\pi_d \pi_{p}(-1)^l,
\end{equation}
where $J$, $\pi$, $J_d$, $\pi_d$, $J_{p}$ and $\pi_{p}$ are spin and parity values of the parent, daughter and emitted proton, respectively.

\begin{figure*}[!htp]
	\centering
	\subfigure[]{
		\label{fig1A}
		\includegraphics[width=0.45\textwidth]{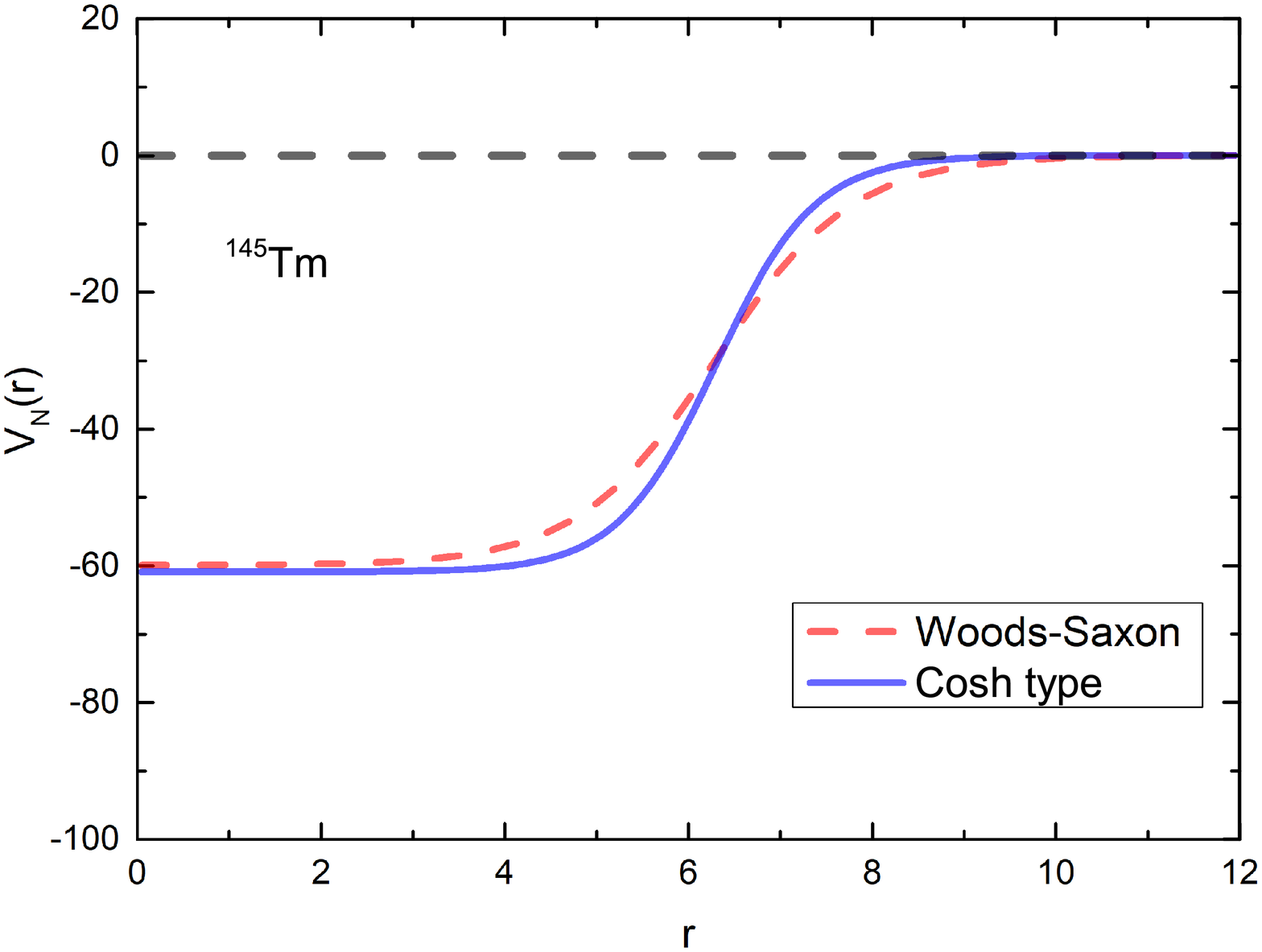}}
	\hspace{1mm}
	\subfigure[]{
		\label{fig1B}
		\includegraphics[width=0.45\textwidth]{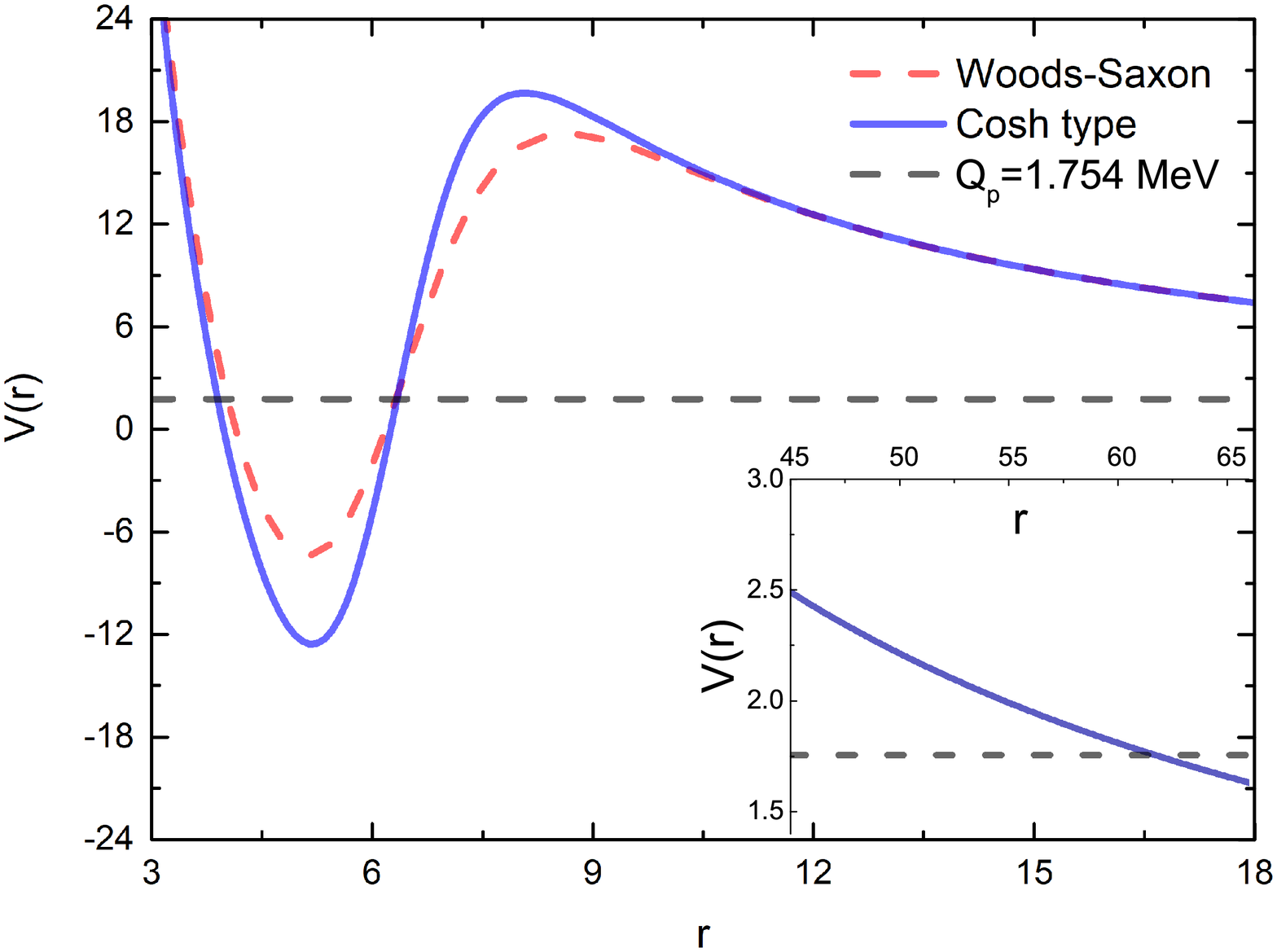}}
	\caption{(color online) The variation of the potential $V(r)$ corresponding to Woods-Saxon and cosh type with separation radius $r$ for nucleus $^{145}$Tm. The part (a) and (b) are about the nuclear and the total potential, respectively.}
	\label{fig1}
\end{figure*}

\section{Results and discussions}
\label{section 3}

In order to extend the cosh type nuclear potential to proton radioactivity, we systematically study the half-lives of 32 spherical proton emitters for $\rm{Z\ge 69}$ with two--potential approach. Firstly, adopting the minimize deviation between the experimental data and calculated results, we determined the parameters of  cosh type nuclear potential $\it i.e.$ the depth $V_0$ and diffuseness $a$ of nuclear potential while
$S_{p}=S_{0}=0.5$ is an approximation taken from Ref. \cite{DELION2006113}. 
 The standard deviation $\sigma$ indicating the deviation between the experimental data and calculated ones can be expressed as
\begin{equation}\label{delta}
\sigma=\sqrt{\frac{1}{N}\sum\limits_{i=1}^N(\rm{log}_{10}T^{\rm{calc},\it i}_{1/2}-\rm{log}_{10}T^{\rm{expt},\it i}_{1/2})^2}.
\end{equation}
By analyzing the experimental proton radioactivity half-lives of 32 spherical proton emitters, we have obtained a set of nuclear potential parameters, $\it i.e.$, $V_0=$ 58.405 MeV and $a=$ 0.537 fm, while the value of renormalization factor $\lambda$  is varied from 0.98 to 1.05 in the whole calculations. The minimize standard deviation is $\sigma=$ 0.297. For an intuitively description of the cosh type nuclear potential, taking nucleus $^{145}$Tm as an example, we plot the comparison of the nuclear and total potential of  cosh type with the ones of Woods-Saxon type in Fig.  \ref{fig1A} and \ref{fig1B}, respectively. In these figures, the red dashed line represents the Woods-Saxon type, and the blue solid line represents the cosh type. Compared with the relatively smooth red dashed line, the turning point of the blue solid line is closer to the surface of the nucleus. Using this set of nuclear potential parameters, we systematically calculate the half-lives of spherical proton emitters for $\rm{Z\ge 69}$ with TPA denoted as log$_{10}T^{\rm{calc1}}_{1/2}$. The detailed results are listed in the seventh column in the Table \ref{tab1}. In this Table, the experimental proton radioactivity half-lives, spin and parity are taken from Refs. \cite{zdeb2016,BLANK2008403,1674-1137-41-3-030001}.  The  released energies $Q_p$ are given by Eq. \ref{q-p} with the mass excess of parent and daughter nuclei taken from Refs. \cite{Huang_2017,Wang_2017} except for $^{144}$Tm, $^{150,151}$Lu, $^{159}$Re, $^{159}$Re$^{m}$, $^{164}$Ir, which are taken from Ref. \cite{BLANK2008403}.

\begin{table*}[!htb]
\centering
\caption{Calculations of proton radioactivity half-lives for spherical proton emitters using different methods and formation probability of proton radioactivity using Eq. \ref{spl}. The calculations displayed as log$_{10}T^{\rm{calc2}}_{1/2}$ and log$_{10}T^{\rm{UDLP2}}_{1/2}$ are performed with considering formation probability of proton radioactivity.}
\label{tab1}
\footnotesize
\begin{threeparttable}
\begin{tabular}{lcccccccccc}
\hline\noalign{\smallskip}
{Nucleus}& $Q_{p}$  & ${l}$ &S$_p^{\rm{expt}}  $ & S$_p^{\rm{calc}}$ & & & &log$_{10}T_{1/2} $\ (\rm{s}) \\
\cline{6-11}
 &(\rm{MeV})& & & &{\rm{expt}}&{\rm{calc1}}  &{\rm{calc2}} & {\rm{UDLP1}} &{\rm{UDLP2}} & {\rm{Gamow-like}}  \\
\noalign{\smallskip}\hline\noalign{\smallskip}
$^{144}\mathrm{Tm   }$&1.725&     5&   1.349       &  1.374    & $-$   5.569&  $-$   5.138     &  $-$     5.577    & $-$   4.691&  $-$   5.218 &  $-$   4.965   \\
$^{145}\mathrm{Tm   }$&1.754&     5&   0.720       &  1.280    & $-$   5.499&  $-$   5.340     &  $-$     5.749    & $-$   4.871&  $-$   5.398 &  $-$   5.164   \\
$^{146}\mathrm{Tm   }$&0.904&     0&   0.413       &  0.505    & $-$   0.810&  $-$   0.893     &  $-$     0.897    & $-$   0.610&  $-$   1.343 &  $-$   0.773   \\
$^{146}\mathrm{Tm^m }$&1.214&     5&   0.969       &  1.192    & $-$   1.125&  $-$   0.837     &  $-$     1.215    & $-$   0.896&  $-$   1.240 &  $-$   0.776   \\
$^{147}\mathrm{Tm   }$&1.133&     2&   0.811       &  0.761    & $-$   3.444&  $-$   3.234     &  $-$     3.416    & $-$   2.859&  $-$   3.603 &  $-$   3.051   \\
$^{147}\mathrm{Tm^m }$&1.072&     5&   0.988       &  1.111    & \ \ \ 0.573& \ \ \  0.869     & \ \ \    0.522    & \ \ \ 0.614& \ \ \  0.344 & \ \ \  0.874   \\
$^{150}\mathrm{Lu^m }$&1.305&     2&   0.381       &  0.611    & $-$   4.398&  $-$   4.516     &  $-$     4.603    & $-$   4.050&  $-$   4.697 &  $-$   4.367   \\
$^{150}\mathrm{Lu   }$&1.283&     5&   0.698       &  0.902    & $-$   1.194&  $-$   1.035     &  $-$     1.291    & $-$   1.113&  $-$   1.316 &  $-$   1.044   \\
$^{151}\mathrm{Lu^m }$&1.335&     2&   0.460       &  0.568    & $-$   4.783&  $-$   4.819     &  $-$     4.875    & $-$   4.327&  $-$   4.978 &  $-$   4.662   \\
$^{151}\mathrm{Lu   }$&1.255&     5&   0.701       &  0.841    & $-$   0.896&  $-$   0.749     &  $-$     0.976    & $-$   0.863&  $-$   1.046 &  $-$   0.767   \\
$^{155}\mathrm{Ta   }$&1.466&     5&   0.800       &  0.640    & $-$   2.495&  $-$   2.291     &  $-$     2.398    & $-$   2.269&  $-$   2.365 &  $-$   2.267   \\
$^{156}\mathrm{Ta   }$&1.036&     2&   0.727       &  0.397    & $-$   0.828&  $-$   0.665     &  $-$     0.565    & $-$   0.624&  $-$   0.949 &  $-$   0.649   \\
$^{156}\mathrm{Ta^m }$&1.126&     5&   1.295       &  0.598    &\ \ \  0.924&\ \ \   1.337     &\ \ \     1.259    &\ \ \  0.947&\ \ \   1.000 &\ \ \   1.248   \\
$^{157}\mathrm{Ta   }$&0.956&     0&   0.792       &  0.552    & $-$   0.529&  $-$   0.329     &  $-$     0.372    & $-$   0.188&  $-$   0.562 &  $-$   0.305   \\
$^{159}\mathrm{Re   }$&1.816&     5&   0.749       &  0.489    & $-$   4.678&  $-$   4.503     &  $-$     4.493    & $-$   4.268&  $-$   4.304 &  $-$   4.428   \\
$^{159}\mathrm{Re^m }$&1.831&     5&   0.622       &  0.489    & $-$   4.695&  $-$   4.600     &  $-$     4.591    & $-$   4.355&  $-$   4.395 &  $-$   4.524   \\
$^{160}\mathrm{Re   }$&1.286&     2&   0.466       &  0.300    & $-$   3.164&  $-$   3.194     &  $-$     2.972    & $-$   2.939&  $-$   3.212 &  $-$   3.090   \\
$^{161}\mathrm{Re   }$&1.216&     0&   0.589       &  0.570    & $-$   3.357&  $-$   3.286     &  $-$     3.342    & $-$   2.895&  $-$   3.234 &  $-$   3.152   \\
$^{161}\mathrm{Re^m }$&1.336&     5&   0.643       &  0.428    & $-$   0.680&  $-$   0.570     &  $-$     0.503    & $-$   0.789&  $-$   0.657 &  $-$   0.601   \\
$^{164}\mathrm{Ir   }$&1.844&     5&   0.239       &  0.352    & $-$   3.959&  $-$   4.279     &  $-$     4.126    & $-$   4.114&  $-$   3.989 &  $-$   4.213   \\
$^{165}\mathrm{Ir^m }$&1.737&     5&   0.378       &  0.330    & $-$   3.430&  $-$   3.551     &  $-$     3.370    & $-$   3.472&  $-$   3.310 &  $-$   3.502   \\
$^{166}\mathrm{Ir   }$&1.177&     2&   0.164       &  0.198    & $-$   0.842&  $-$   1.325     &  $-$     0.924    & $-$   1.303&  $-$   1.336 &  $-$   1.294   \\
$^{166}\mathrm{Ir^m }$&1.347&     5&   0.417       &  0.309    & $-$   0.091&  $-$   0.170     &\ \ \     0.040    & $-$   0.475&  $-$   0.173 &  $-$   0.215   \\
$^{167}\mathrm{Ir   }$&1.087&     0&   0.685       &  0.597    & $-$   1.128&  $-$   0.991     &  $-$     1.068    & $-$   0.865&  $-$   0.945 &  $-$   0.940   \\
$^{167}\mathrm{Ir^m }$&1.262&     5&   0.479       &  0.290    & \ \ \ 0.778&\ \ \   0.759     &\ \ \     0.997    & \ \ \ 0.348&\ \ \   0.693 &\ \ \   0.683   \\
$^{170}\mathrm{Au   }$&1.487&     2&   0.119       &  0.151    & $-$   3.487&  $-$   4.111     &  $-$     3.592    & $-$   3.845&  $-$   3.839 &  $-$   3.984   \\
$^{170}\mathrm{Au^m }$&1.767&     5&   0.212       &  0.239    & $-$   2.975&  $-$   3.347     &  $-$     3.026    & $-$   3.333&  $-$   3.012 &  $-$   3.307   \\
$^{171}\mathrm{Au   }$&1.464&     0&   0.402       &  0.615    & $-$   4.652&  $-$   4.746     &  $-$     4.836    & $-$   4.298&  $-$   4.378 &  $-$   4.569   \\
$^{171}\mathrm{Au^m }$&1.718&     5&   0.195       &  0.224    & $-$   2.587&  $-$   2.997     &  $-$     2.648    & $-$   3.026&  $-$   2.683 &  $-$   2.966   \\
$^{176}\mathrm{Tl   }$&1.278&     0&   0.476       &  0.639    & $-$   2.208&  $-$   2.229     &  $-$     2.335    & $-$   2.059&  $-$   1.882 &  $-$   2.133   \\
$^{177}\mathrm{Tl   }$&1.172&     0&   0.946       &  0.643    & $-$   1.178&  $-$   0.901     &  $-$     1.011    & $-$   0.863&  $-$   0.626 &  $-$   0.855   \\
$^{177}\mathrm{Tl^m }$&1.979&     5&   0.060       &  0.154    & $-$   3.459&  $-$   4.378     &  $-$     3.866    & $-$   4.294&  $-$   3.848 &  $-$   4.317   \\

\noalign{\smallskip}\hline
\end{tabular}
\end{threeparttable}
\end{table*}

\begin{figure*}[htp!]
\centering
\includegraphics[width=17cm]{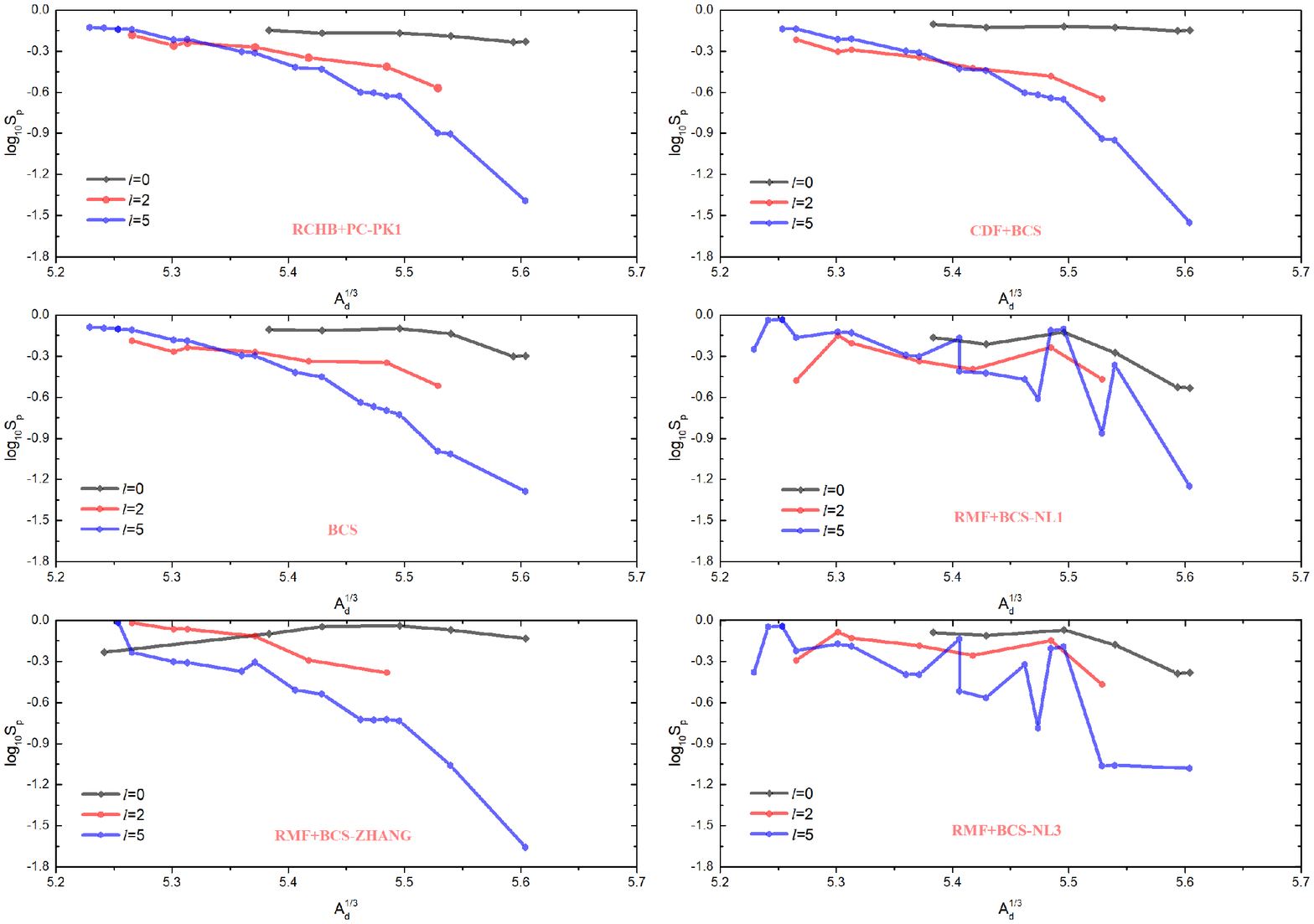}
\caption{\label{fig2}(color online) Formation probability of proton radioactivity in different semi-microscopic methods  as a function of $A_d^{1/3}$. The methods include BCS-Lim \cite{PhysRevC.93.014314}, RMF+BCS-Zhang \cite{Zhang_2010}, CDF+BCS \cite{PhysRevC.90.054326}, RCHB+PC-PL1 \cite{PhysRevC.93.014314}, RMF+BCS-NL1 \cite{Soylu_2021} and RMF+BCS -NL3 \cite{Soylu_2021}.}
\end{figure*}

\begin{figure}[!]
\centering
\includegraphics[width=8.5cm]{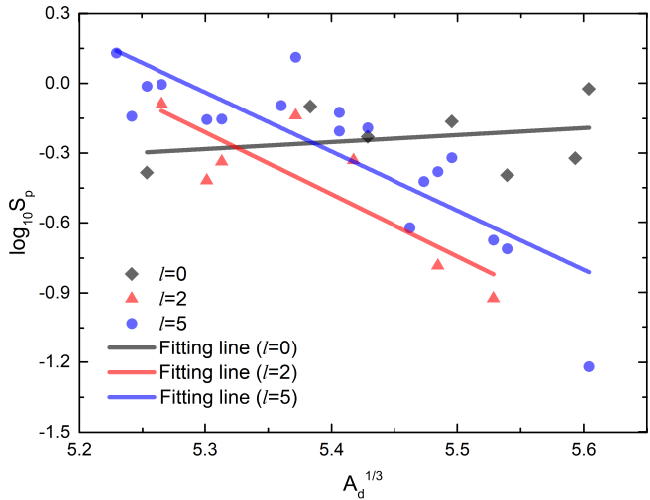}
\caption{\label{fig3}(color online) The linear relationship between the formation probability of proton radioactivity $S_p^{\rm expt}$ and $A_d^{1/3}$.}
\end{figure}

\begin{figure}[!]
\centering
\includegraphics[width=8.5cm]{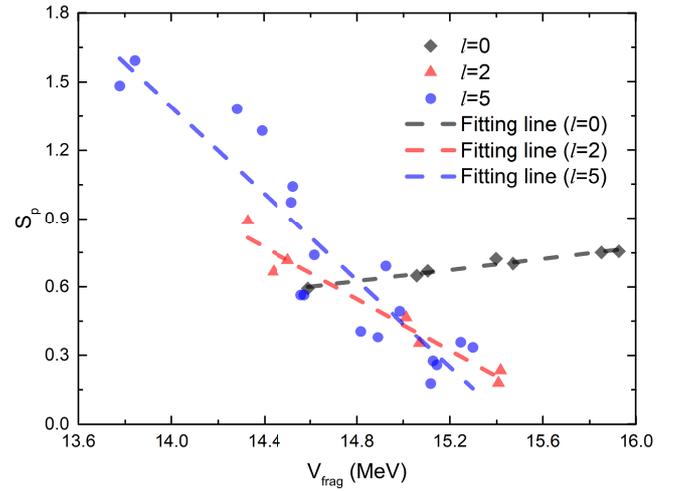}
\caption{\label{figvfrag}(color online) The linear relationship between the formation probability of proton radioactivity $S_p^{\rm calc}$ calculated by Eq. \ref{spl} and the fragmentation potential $V_{\rm frag}$.}
\end{figure}

The formation probability of proton radioactivity containing a lot of nuclear structure properties is also called by spectroscopy factor, which plays an important role in the calculation of half--life. The experimental formation probability of proton radioactivity can be extracted from ratios of calculated proton radioactivity half-life $T_{1/2}^{\rm calc}$ to experimental data $T_{1/2}^{\rm expt}$, which is defined as $S_p^{\rm expt}=S_{0}T_{1/2}^{\rm calc}/T_{1/2}^{\rm expt}$. Both the semi-microscopic method and the phenomenological method are used to calculate the probability of formation of proton radioactivity \cite{Zhang_2010,PhysRevC.79.054330,Qian201668,Soylu_2021,PhysRevC.90.054326,PhysRevC.93.014314,PhysRevC.85.011303,QI2021136373,PhysRevC.103.054325,QI2019214}. In 2012,  Qi ${\it et\ al.}$ \cite{PhysRevC.85.011303} proposed a universal decay law for proton emission (UDLP). Meanwhile they studied the  correlation between formation probability of proton radioactivity  and nuclear structure properties. In UDLP, the logarithm of proton radioactivity half-life can be expressed as
\begin{equation}
\rm{log}_{10}T_{1/2}={\it A}\chi'+{\it B}\rho'+{\it C}\frac{l(l+1)}{\rho'}+{\it D},
\end{equation}
where $A$, $B$, $C$ and $D$ are adjustable parameters, and their different sets of values are taken from Ref. \cite{PhysRevC.85.011303}. $\chi'=Z_d\sqrt{A_d/((A_d+1)Q_p)}$, $\rho'=\sqrt{A_dZ_d(A_d^{1/3}+1)/(A_d+1)}$. In their study, there is a similar linear relationship between the formation probability of proton radioactivity and $\rho'$ while $\rho' \approx \rho=\mu\nu R/\hbar$ with $\nu$ being the classic assault frequency. In this work, we study the relationship between the formation probability of proton radioactivity and $A_d^{1/3}$ which is one of the main nuclear structural properties in $\rho'$ or $\rho$. 
We collect some recent results of formation probability calculated by semi-microscopic methods and discuss their direct relationship with the properties of the nuclear structure by phenomenological means. The methods including BCS-Lim \cite{PhysRevC.93.014314}, RMF+BCS-Zhang \cite{Zhang_2010}, CDF+BCS \cite{PhysRevC.90.054326}, RCHB+PC-PL1 \cite{PhysRevC.93.014314}, RMF+BCS-NL1 \cite{Soylu_2021} and RMF+BCS -NL3 \cite{Soylu_2021} are plotted in Fig. \ref{fig2}. In this figure,  each kind of data show a good linear relationship between the formation probability of proton radioactivity and $A_d^{1/3}$ with the same orbital angular  momentum $l$. On this basis, 
adopting the experimental formation probability $S_p^{\rm expt}$ listed in the fourth column in the Table \ref{tab1},  we plot $S_p^{\rm expt}$ as a function of $A_d^{1/3}$ in Fig. \ref{fig3} . In this figure, with the orbital angular momentum $l$ remaining the same, there is an obvious linear relationship between $S_p^{\rm expt}$ and $A_d^{1/3}$. Therefore, we propose a simple linear relationship to evaluate the formation probability of proton radioactivity as
\begin{equation}\label{spl}
\rm{log}_{10}S_{p}={\it a_l}A_d^{1/3}+{\it b_l},
\end{equation}
where $a_l$ and $b_l$ are adjustable parameters with $l=0,\  2,\  5$. The detailed results of these parameters are listed in Table \ref{tab2}. The calculations of  formation probability of proton radioactivity denoted as $S_p^{\rm calc}$ are listed in the fifth column in Table \ref{tab1}. By considering the formation probability obtained by Eq. \ref{spl}, we use the TPA to evaluate the proton radioactivity half--lives denoted as log$_{10}T^{\rm{calc2}}_{1/2}$. Adopting the set of parameters for $N\geq 75$ proposed by Qi ${\it et\ al.}$ while the formation probability is considered, we use UDLP to evaluate the proton radioactivity half--lives denoted as log$_{10}T^{\rm{UDLP2}}_{1/2}$. The both results are listed in the Table \ref{tab1}. In addition, Delion noted that \cite{PhysRevC.80.024310} connecting the formation probability (spectroscopic factor) with the mass of the emitted proton can be nicely explained in terms of the fragmentation potential, which is given by the difference between the Coulomb barrier and the $Q_p$ value as
 \begin{equation}
V_{\rm frag}=\frac{Z_de^2}{R}-Q_p.
 \end{equation}
As a verification, we plot the formation probability $S_p^{\rm calc}$ calculated by Eq. \ref{spl}  versus with the fragmentation potential $V_{\rm frag}$ in Fig. \ref{figvfrag}. In this figure, classified by orbital angular momentum $l$ same as in Fig. \ref{fig2}, there is an obvious linear relationship between $S_p^{\rm calc}$ and $V_{\rm frag}$. Therefore,  there is a correlation between formation probability of the proton radioactivity  and the nuclear structure properties. It indicates that the formation probability of proton radioactivity can be simply described by a formula of $A_d^{1/3}$.

\begin{table}[]
\centering
\caption{The adjustable parameters value of $a_l$ and $b_l$ with $l=0,\  2,\  5$.}
\label{tab2}
\begin{tabular}{lccc}
\hline\noalign{\smallskip}
Para.  & $l=0$\   &$l=2$\     &$l=5$\    \\
\noalign{\smallskip}\hline\noalign{\smallskip}
$a_l$ &\ \   0.301 & $-$2.665 &  $-$2.537 \\  
$b_l$      & $-$1.877  &\ \    13.912 &\ \     13.405 \\ 
\noalign{\smallskip}\hline
\end{tabular}
\end{table}

\begin{figure}[!]
\centering
\includegraphics[width=8.5cm]{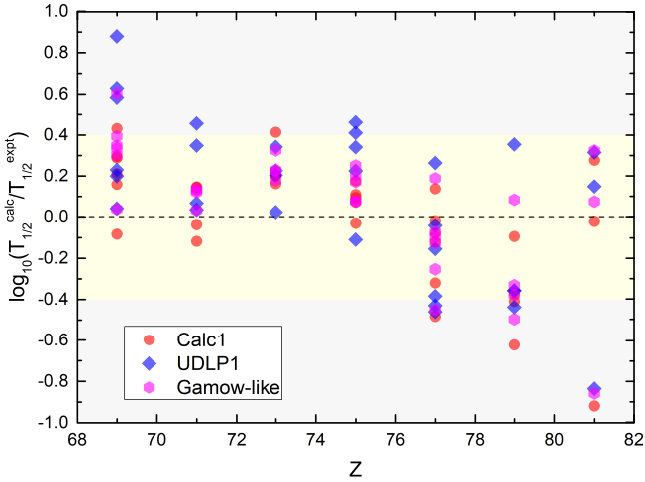}
\caption{\label{fig4}(color online) Decimal logarithm deviations between the experimental data of proton radioactivity half-lives and calculations. The circles, rhombuses and hexagons refer to results obtained by our method, UDLP and Gamow--like model, denoted as Calc1, UPLP1 and Gamow--like, respectively. These results are performed without considering formation probability of proton radioactivity.}
\end{figure}

\begin{figure}[!]
\centering
\includegraphics[width=8.5cm]{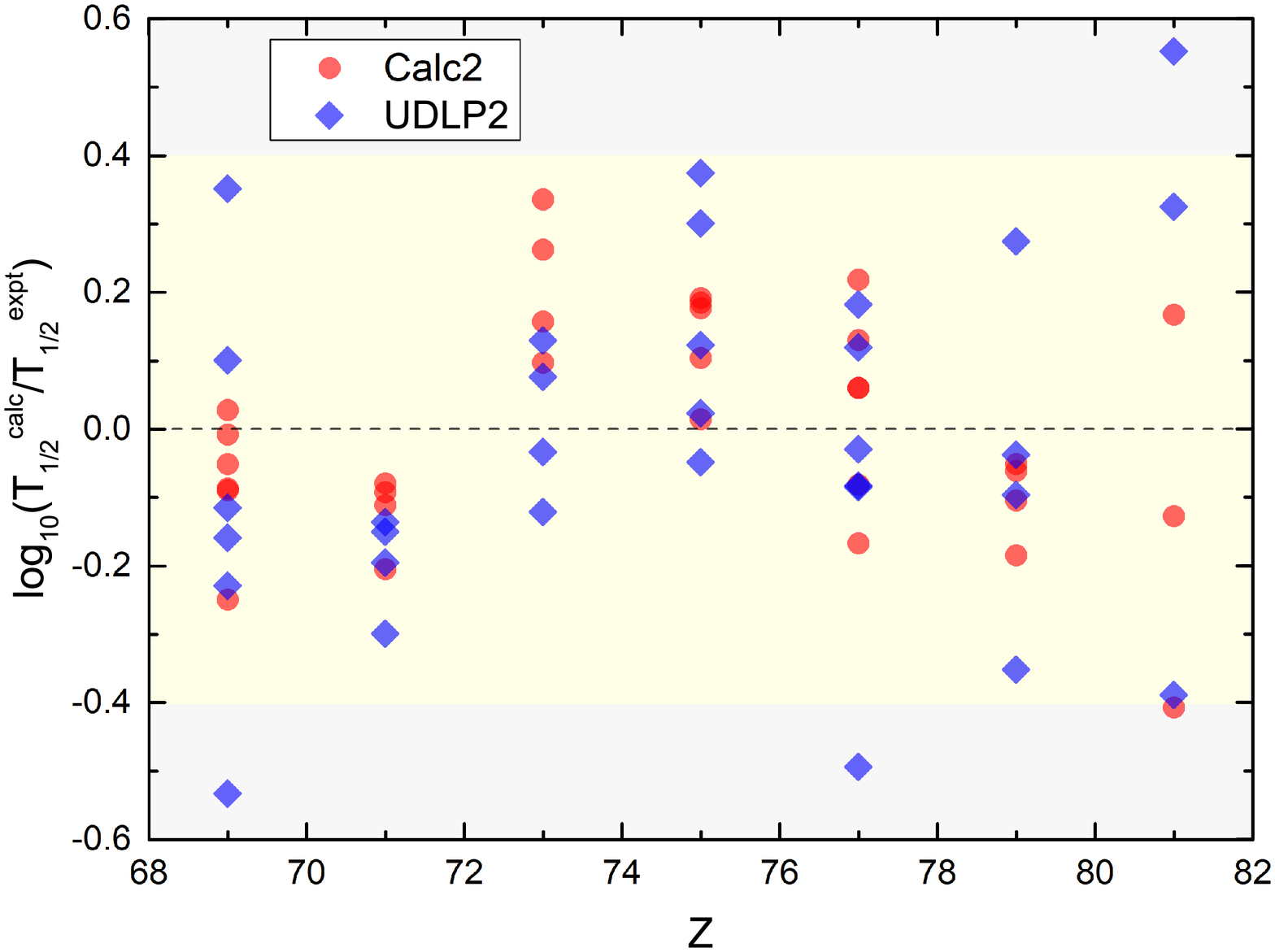}
\caption{\label{fig5}(color online) Decimal logarithm deviations between the experimental data of proton radioactivity half-lives and calculations. The circles and rhombuses refer to results obtained by our method and UDLP, denoted as Calc2 and UPLP2, respectively. These results are performed with considering formation probability of proton radioactivity.}
\end{figure}

In order to visually compare the calculated half-lives of proton radioactivity in the Table \ref{tab1}, we plot the logarithmic differences between log$_{10}T^{\rm{calci}}_{1/2}$  and log$_{10}T^{\rm{expt}}_{1/2}$  versus with proton number of parent nuclei in Fig. \ref{fig4} and \ref{fig5}. In the Fig. \ref{fig4}, the calculated half-lives are evaluated by the present method, UDLP \cite{PhysRevC.85.011303} and Gamow-like model \cite{Chen_2019} without considering the formation probability of proton radioactivity. The results in the figure denoted as the red circle, purple upper-triangle and blue triangle are generally in the range of $\pm 0.4$, and all in the range of $\pm 1.0$, corresponding to the ratio $T^{\rm{calc}}_{1/2}$ /$T^{\rm{expt}}_{1/2}$ within a factor of about 0.4--2.5. In the Fig. \ref{fig5},  the calculated half-lives are evaluated by the present method and UDLP  with considering the formation probability of proton radioactivity. The results in the figure are generally in the range of $\pm 0.4$, and all in the range of $\pm 0.6$. In addition, by using Eq. \ref{V_N}, the standard deviations $\sigma$ of these 5 cases from the Fig. \ref{fig4} and \ref{fig5} are listed in the Table {\ref{tab3}}. In this table, the results of the case in figure \ref{fig4} show that $\sigma_{\rm{calc1}}$= 0.297, $\sigma_{\rm{UDLP1}}$=0.385 and $\sigma_{\rm{G-L}}$=0.304. It indicates that  the ratio $T^{\rm{calc}}_{1/2}$ /$T^{\rm{expt}}_{1/2}$ is in a factor of about 2.01--2.43 without considering the formation probability of proton radioactivity. The results of the case in Fig. \ref{fig5} show that $\sigma_{\rm{calc2}}$=0.163 and $\sigma_{\rm{UDLP2}}$=0.253.  It indicates that  the $\sigma$ in contrast to the case of Fig. \ref{fig4} is improved. Compared with the previous calculations, with considering formation probability, our present work displayed as ${\rm{calc2}}$ significant improves by $\frac{0.297-0.163}{0.297}=45.1\%$ and the UDLP displayed as ${\rm{UDLP2}}$ significant improves by $\frac{0.385-0.253}{0.385}=34.3\%$. It indicates that the accuracy of the calculated half--life can be further improved with considering the formation probability. Moreover, with the orbital angular momentum $l$ remaining the same, there is an obvious linear relationship between $S_p^{\rm expt}$ and $A_d^{1/3}$. The formation probability of proton radioactivity can be simply described by a formula of $A_d^{1/3}$.

\begin{table}[]
\centering
\caption{The standard deviations $\sigma$ of this work, Gamow-like and UDLP in different cases.}
\label{tab3}
\begin{tabular}{lccccc}
\hline\noalign{\smallskip}
Type  &  Calc1   & Calc2     &UDLP1  &UDLP2 &Gamow--like   \\
\noalign{\smallskip}\hline\noalign{\smallskip}
$\sigma$&  0.297 & 0.163 &  0.385 &  0.253 &  0.304\\
\noalign{\smallskip}\hline
\end{tabular}
\end{table}

\section{Summary}
\label{section 4}
In summary, we systematically study the half--lives of proton radioactivity for $\rm{Z\ge 69}$ spherical proton emitters based on two--potential approach with a cosh type potential.  By fitting 32 experimental data,  the parameters of the depth and diffuseness for cosh type nuclear potential are determined as $V_0=$ 58.405 MeV and $a=$ 0.537 fm, respectively. In addition, for studying the formation probability of proton radioactivity, we propose a simple analytic expression for describing the relationship between the formation probability extracted from the
experimental half-life of proton radioactivity and the quantity $A_d^{1/3}$. The standard deviation obtained with respect to the experimental half-lives suggested that the present systematic study can be applied successfully to study proton radioactivity. \\

\noindent
This work is supported in part by National Natural Science Foundation of China (Grant No. 11205083 and No. 11975132 ), the construct program of the key discipline in Hunan province, the Innovation Group of Nuclear and Particle Physics in USC, the Shandong Province Natural Science Foundation, China (Grant No. ZR2019YQ01), the Opening Project of Cooperative Innovation Center for Nuclear Fuel Cycle Technology and Equipment, University of South China (2019KFZ10), and the Hunan Provincial Innovation Foundation For Postgraduate (Grant No. CX20190714).

%

\begin{thebibliography}{}
%
%

\bibliographystyle{unsrt}

\bibitem{JACKSON1970281}K. Jackson, C. Cardinal, H. Evans, N. Jelley, and J. Cerny, {\it Phys. Lett. B} \textbf{33}, 281 (1970)
\bibitem{CERNY1970284}J. Cerny, J. Esterl, R. Gough, and R. Sextro, {\it Phys. Lett. B} \textbf{33}, 284 (1970)

\bibitem{Hofmann1982}S. Hofmann, W. Reisdorf, G. M{\"u}nzenberg, F. P. He{\ss}berger, J. R. H. Schneider, and P. Armbruster, {\it Z. Phys. A} \textbf{305}, 111 (1982)
\bibitem{Klepper1982}O. Klepper, T. Batsch, S. Hofmann, R. Kirchner, W. Kurcewicz, W. Reisdorf, E. Roeckl, D. Schardt, and G. Nyman, {\it Z. Phys. A} \textbf{305}, 125 (1982)
\bibitem{Faestermann1984}T. Faestermann, A. Gillitzer, K. Hartel, P. Kienle, and E. Nolte, {\it Phys. Lett. B} \textbf{137}, 23 (1984)
\bibitem{SONZOGNI20021}A. Sonzogni, {\it Nucl. Data Sheets} \textbf{95}, 1 (2002)
\bibitem{PhysRevLett.96.072501}D. S. Delion, R. J. Liotta, and R. Wyss, {\it Phys. Rev. Lett.} \textbf{96}, 072501 (2006)
\bibitem{BLANK2008403}B. Blank and M. Borge,  {\it Prog. Part. Nucl. Phys.}\textbf{ 60}, 403 (2008)
\bibitem{Zhang_2010}H. F. Zhang, Y. J. Wang, J. M. Dong, J. Q. Li, and W. Scheid,  {\it J. Phys. G: Nucl. Part. Phys.} \textbf{37}, 085107 (2010)
\bibitem{PhysRevC.96.034619}K. P. Santhosh and I. Sukumaran,   {\it Phys. Rev. C} \textbf{96}, 034619 (2017)
\bibitem{Budaca2017}R. Budaca and A. I. Budaca,  {\it Eur. Phys. J. A} \textbf{53}, 160 (2017)
\bibitem{1674-1137-41-3-030001}G. Audi, F. G. Kondev, M. Wang, W. Huang, and S. Naimi,  {\it Chin. Phys. C} \textbf{41}, 030001 (2017) 
\bibitem{Chen_2019} J. L. Chen, X. H. Li, J. H. Cheng, J. G. Deng, and X. J. Wu,  {\it J. Phys. G: Nucl. Part. Phys.} \textbf{46}, 065107 (2019)
\bibitem{KARNY200852}M. Karny \textit{et al.}, {\it Phys. Lett. B} \textbf{664}, 52 (2008) 
\bibitem{PhysRevC.72.051601}D. N. Basu, P. R. Chowdhury, and C. Samanta,   {\it Phys. Rev. C} \textbf{72}, 051601 (2005) 
\bibitem{BHATTACHARYA2007263}M. Bhattacharya and G. Gangopadhyay,   {\it Phys. Lett. B} \textbf{651}, 263 (2007) 
\bibitem{QIAN-Yi-Bin-72301}Y. B. Qian, Z. Z. Ren, and D. D. Ni,  {\it Chin. Phys. Lett.} \textbf{27}, 072301 (2010) 
\bibitem{QIAN-Yi-Bin**-112301}Y. B. Qian \textit{et al.}, {\it Chin. Phys. Lett.} \textbf{27}, 112301 (2010) 
\bibitem{PhysRevC.79.054330}J. M. Dong, H. F. Zhang, and G. Royer,   {\it Phys. Rev. C} \textbf{79}, 054330 (2009) 
\bibitem{yzwang2017}Y. Z. Wang, J. P. Cui, Y. L. Zhang, S. Zhang, and J. Z. Gu,   {\it Phys. Rev. C} \textbf{95}, 014302 (2017) 
\bibitem{PhysRevC.71.014603}M. Balasubramaniam and N. Arunachalam,   {\it Phys. Rev. C} \textbf{71}, 014603 (2005) 
\bibitem{1674-1137-34-2-005}J. M. Dong, H. F. Zhang, W. Zuo, and J. Q. Li,  {\it Chin. Phys. C} \textbf{34}, 182 (2010) 
\bibitem{ZhangGL2013}C. Guo, G. Zhang, and X. Le,   {\it Nucl. Phys. A} \textbf{897}, 54 (2013) 
\bibitem{ZhangGL2014}C. Guo and G. Zhang,   {\it Eur. Phys. J. A} \textbf{50}, 187 (2014) 
\bibitem{zdeb2016}A. Zdeb, M. Warda, C. M. Petrache, and K. Pomorski,  {\it Eur. Phys. J. A} \textbf{52}, 323 (2016) 
\bibitem{CHENG2020121717}J. H. Cheng, J. L. Chen, J. G. Deng, X. H. Li, Z. Zhang, and P. C. Chu,   {\it Nucl. Phys. A} \textbf{997}, 121717 (2020) 
\bibitem{PhysRevC.45.2247}B. Buck, A. C. Merchant, and S. M. Perez,   {\it Phys. Rev. C} \textbf{45}, 2247 (1992)
\bibitem{PhysRevLett.72.1326}B. Buck, A. C. Merchant, and S. M. Perez,   {\it Phys. Rev. Lett.} \textbf{72}, 1326 (1994)
\bibitem{PhysRevC.93.034316}X. D. Sun, P. Guo, and X. H. Li,   {\it Phys. Rev. C} \textbf{93}, 034316 (2016)
\bibitem{PhysRevC.94.024338}X. D. Sun, P. Guo, and X. H. Li,   {\it Phys. Rev. C} \textbf{94}, 024338 (2016)
\bibitem{PhysRevC.95.044303}X. D. Sun, J. G. Deng, D. Xiang, P. Guo, and X. H. Li,  {\it Phys. Rev. C} \textbf{95}, 044303 (2017)
\bibitem{Deng_2018}J. G. Deng, J. C. Zhao, J. L. Chen, X. J. Wu, and X. H. Li,   {\it Chin. Phys. C} \textbf{42}, 044102 (2018)
\bibitem{PhysRevLett.59.262}S. A. Gurvitz and G. Kalbermann,   {\it Phys. Rev. Lett.} \textbf{59}, 262-265 (1987)
\bibitem{Zhang072301}H. F. Zhang \textit{et al.},  {\it Chin. Phys. Lett.} \textbf{26}, 072301 (2009)

\bibitem{PhysRevC.56.1762}S. \AA{}berg, P. B. Semmes, and W. Nazarewicz, {\it Phys. Rev. C} \textbf{56}, 1762 (1997)
\bibitem{DELION2006113}D. Delion, R. Liotta, and R. Wyss,   {\it Phys Rep}, \textbf{80}, 424 (2006)
\bibitem{Qian201668}Y. Qian and Z. Ren, {\it Eur. Phys. J. A} \textbf{52}, 68 (2016)
\bibitem{Soylu_2021}A. Soylu, F. Koyuncu, G. Gangopadhyay, V. Dehghani, and S. A. Alavi, {\it Chin. Phys. C} \textbf{45}, 044108 (2021)
\bibitem{PhysRevC.90.054326}Q. Zhao, J. M. Dong, J. L. Song, and W. H. Long, {\it Phys. Rev. C} \textbf{90}, 054326 (2014)
\bibitem{PhysRevC.93.014314}Y. Lim, X. Xia, and Y. Kim, {\it Phys. Rev. C} \textbf{93}, 014314 (2016)
\bibitem{PhysRevC.85.011303}C. Qi, D. S. Delion, R. J. Liotta, and R. Wyss,   {\it Phys. Rev. C} \textbf{85}, 011303(R) (2012)
\bibitem{QI2019214}C. Qi, R. Liotta, and R. Wyss, {\it Prog. Part. Nucl. Phys.} \textbf{105}, 214 (2019)
\bibitem{QI2021136373}C. Qi, R. J. Liotta, and R. Wyss, {\it Phys. Lett. B} \textbf{818}, 136373 (2021)
\bibitem{PhysRevC.103.054325}D. S. Delion and A. Dumitrescu, {\it Phys. Rev. C} \textbf{103}, 054325 (2021)
\bibitem{Huang_2017}W. Huang, G. Audi, M. Wang, F. G. Kondev, S. Naimi, and X. Xu,   {\it Chin. Phys. C} \textbf{41}, 030002 (2017)
\bibitem{Wang_2017}M. Wang, G. Audi, F. G. Kondev, W. Huang, S. Naimi, and X. Xu,  {\it Chin. Phys. C} \textbf{41}, 030003 (2017)
\bibitem{PhysRevC.72.064613}V. Y. Denisov and H. Ikezoe,   {\it Phys. Rev. C} \textbf{72}, 064613 (2005)
\bibitem{Kelkar064605}N.G. Kelkar, H.M. Castaneda,  {\it Phys. Rev. C} \textbf{76}, 064605 (2007)
\bibitem{Buck1688}B. Buck, A. C. Merchant, and S. M. Perez,   {\it Phys. Rev. C} \textbf{45}, 1688 (1992)
\bibitem{doi:10.1063/1.531270} J. J. Morehead,  {\it J. Math. Phys.} \textbf{36}, 5431 (1995)
\bibitem{HUANG1976243}K. N. Huang, M. Aoyagi, M. H. Chen, B. Crasemann, and H. Mark,   {\it At. Data Nucl. Data Tables} \textbf{18}, 243 (1976)
\bibitem{PhysRevC.80.024310}D. S. Delion,   {\it Phys. Rev. C} \textbf{80}, 024310 (2009)





\end{thebibliography}
%

\end{document}